\documentclass[10pt]{iopart}
\usepackage{fleqn,epsfig,iopams,amssymb,amsbsy,xspace}  
\begin{document}
     \newcommand{\pathstrange}{/home/rafelski/figure/}
     \newcommand{\pathmarc}{/usr1/rafelski/figure/}
     \newcommand{\pathbook}{/usr1/rafelski/book/figures/}
     \newcommand{\pathtransp}{/usr1/rafelski/paper/transp/}
     \newcommand{\pathlaptop}{/home/rafelski/figures/}
     \newcommand{\pathlapbook}{/home/rafelski/book/qgp/figures/}
     \newcommand{\pathletes}{/users/lpthe/jletes/bookraf/figures/}
     \newcommand{\pathjussieu}{/users/visit/rafelski/figures/}
     \newcommand{\pathnow}{}
\renewcommand{\topfraction}{.99}\renewcommand{\textfraction}{.0}
\newcommand{\myfig}[7]{%
\begin{figure}[#7]
\vskip #5cm	\centerline{\hspace*{#1cm}
\epsfig{width=#2cm,angle=#4,figure=\pathnow #3.ps}
	}\vskip #6cm
                    }
\newcommand{\capt}[3]%
{\caption[{#2}]{\label{Fig:#1}#3}
}
\newcommand{\rf}[1]{Fig.\,\ref{Fig:#1}%
}
\def\qgp{quark--gluon plasma\xspace}
\def\QGP{\qgp}

\newcommand{\ress}[1]{subsection~\ref{ssec:#1}%
}
\newcommand{\ressa}[2]{subsections~\ref{ssec:#1},~\ref{ssec:#2}%
}
\newcommand{\ressc}[2]{subsections~\ref{ssec:#1}--\ref{ssec:#2}%
}
\newcommand{\ressb}[3]{section~%
\ref{ssec:#1},~\ref{ssec:#2},~\ref{ssec:#3}%
}
\newcommand{\pref}[1]{on page~\pageref{eq:#1}} 
\newcommand{\req}[1]{Eq.\,(\ref{eq:#1})%
}
\newcommand{\reqp}[1]{Eq.\,(\ref{eq:#1}) on page~\pageref{eq:#1}%
}
\newcommand{\reqa}[2]{Eqs.\,(\ref{eq:#1}, \ref{eq:#2})%
}
\newcommand{\reqc}[2]{Eqs.\,(\ref{eq:#1}--\ref{eq:#2})%
}
\newcommand{\reqb}[3]{Eqs.\,(\ref{eq:#1}, \ref{eq:#2}, \ref{eq:#3})%
}
\newcommand{\beql}[1]{
	\begin{equation} \label{eq:#1}}

\newcommand{\beqarl}[1]{
	\begin{eqnarray} \label{eq:#1} }
\newcommand{\eeql}[1]{\label{eq:#1} \end{equation} 
} 
\newcommand{\eeqarl}[1]{\label{eq:#1} \end{eqnarray} 
}
\newcommand{\rt}[1]{table~\ref{Tab:#1}%
}
\def\beq{\begin{equation}}
\def\eeq{\end{equation}}
\def\neeq{\nonumber \eeq}
\def\beqar{\begin{eqnarray}}
\def\eeqar{\end{eqnarray}}
\def\bcite{\cite}
\def\agev{{$A$~GeV}\xspace}
\def\AGeV{\agev}
\newcommand{\captt}[3]%
{\caption[{#2}]{\label{Tab:#1}#3%
}%
}
\newcommand{\lsec}[1]{\label{sec:#1} } 
\newcommand{\res}[1]{section~\ref{sec:#1}%
}
\newcommand{\resa}[2]{sections~\ref{sec:#1},~\ref{sec:#2}%
}
\newcommand{\resc}[2]{sections~\ref{sec:#1}--\ref{sec:#2}%
}
\newcommand{\resb}[3]{section~%
\ref{sec:#1},~\ref{sec:#2},~\ref{sec:#3}%
}
\newcommand{\lssec}[1]{\label{ssec:#1} 
 }

\def\etc{{etc.\xspace}}
\def\ie{{i.e.\xspace}}
\def\eg{{e.g.\xspace}}
\def\etal{{\it et al.\/}}
\def\viz{{viz.\xspace}}
\def\Jpsi{${\mathrm J}\!/\!\,\Psi$\xspace}
\newcommand{\myfigd}[9]{%
\begin{figure}[#9]
\vskip #5cm	\centerline{\hspace*{#1cm}
\psfig{width=#2cm,angle=#4,figure=\pathnow #3.ps}
\hspace*{#7cm}
\psfig{width=#2cm,angle=#4,figure=\pathnow #8.ps}
	}\vskip #6cm
                    }
\hyphenation{re-commend-ed}

\title{Importance 
       of reaction  volume\,in\,hadronic collisions:\\
Canonical  enhancement}

\author{Johann Rafelski$^{\rm\dag}$
 and Jean Letessier$^{\rm\ddag}$ 
}

\address{\dag\ Department of Physics,
University of Arizona, Tucson, AZ 85721,
and\\ CERN-Theory Division, 1211 Geneva 23, Switzerland}

\address{\ddag\ 
Laboratoire de Physique Th\'eorique et Hautes Energies
\\
Universit\'e Paris 7, 2 place Jussieu, F--75251 Cedex 05.
}

\begin{abstract}
We study the canonical flavor enhancement arising from 
exact conservation of strangeness, and charm flavor. Both the
theoretical motivation, and the practical consequences
 are explored. We argue using qualitative theoretical arguments and 
quantitative evaluation, that this proposal to reevaluate
strangeness signature of quark--gluon plasma is not able to explain
the majority of available experimental results.
\end{abstract}


\submitto{\JPG \rm Proceedings of Strange Quark Matter 2001, Frankfurt}


\vskip -10cm \ \hfill CERN-TH/2001-357 \vskip 10cm

\section{Introduction}
The canonical statistical mechanics method
to treat small hadron abundances constrained by
a conserved charge, or flavor, has been invented  in early days 
of statistical hadron production theory \bcite{Mag55}.
It has taken 20 years before is has been applied to the study of small 
p--p reaction systems \cite{Shu74}. In the early days 
of relativistic heavy ion collisions it has been important to 
validate that size of the physical system considered is
allowing  grand canonical statistical method in study of hadron
production \bcite{Raf80}. Today, with the enhancement of 
strangeness production in A--A reactions obtained from
comparison to expectations derived from study 
of p--p, p--Be, p--Pb collision systems,
a claim has arisen that one can reinterpret the strange 
hadron signature of quark--gluon plasma in terms of the so 
called canonical enhancement/suppression \bcite{Ham00c,Red01},
an issue which we address in depth here. 

We will explain the need to amend the grand canonical method in 
\ress{general}, and present 
the intuitive derivation of the canonical constraint in \ress{canon},
where we follow the approach of Ref.\,\bcite{Raf80}. This
can be generalized to more complex systems using the
projection method \bcite{Red80,Tur81}, which we 
demonstrate in \ress{proj}, and use in \ress{multi} to obtain within the
classical Boltzmann limit the suppression factors of
multistrange hadrons \bcite{Ham00c,Red01,Cle91}. This method
can be extended and applied to solve more complex situation, for example 
conservation of several `Abelian' quantum numbers \bcite{Der85,Bec98} 
(such as strangeness, baryon number, electrical charge) and
the problem of particular relevance in this field, the exact conservation
of color: all hadronic states, including QGP  
must be exactly color `neutral' \bcite{Elz83,Elz84}. 

After offering this thorough  theoretical  introduction in \res{ReacVol},  
 we study, in \res{CanSup}, the magnitude of the different 
effects. We evaluate the magnitude of the canonical suppression
in \ress{suppress}. After a rebasing which 
is converting the suppression into enhancement, we evaluate
in \ress{HGQGP} the canonical enhancement effect for both 
the  hadronic phase and the deconfined phase.
In \ress{charm} we show that for charm flavor, 
the canonical equilibrium mechanism is 
leading to a significant disagreement with experimental
constraints on charm production. Considering that there is no reason to 
expect that flavors such as charm or strangeness differ in
any fundamental way,  this shows that both strangeness and charm
yields need to be studied within the realm of  chemical kinetic 
theory \bcite{Raf82b,Let00b}.   

In the closing \res{close}, we
evaluate and  discuss the specific (per participant) A--A yield
enhancement of multistrange baryons and antibaryons,
comparing with the p--p and p--A collision systems, and we show
that the magnitude of the canonical equilibrium 
enhancement for multistrange hadrons is highly sensitive to 
the  strangeness yield of the reference p--A system.

\section{Exact conservation of flavor quantum numbers}      
\lsec{ReacVol}      
\subsection{General considerations}
\lssec{general}
At low reaction energy, or/and in small collision systems
the yield of strangeness in each reaction is rather small, 
less than one pair of quarks 
produced per collision. This occasional pair can 
thermally (momentum distribution) equilibrate with the background of hadrons.
In the discussion of the magnitude of
this yield, we may be tempted to apply methods of grand canonical 
statistical ensemble equilibrium. However, these are  wrong, as in their 
derivation a strong and important assumption is that the number
of particles considered is large. 

The statistical  grand canonical 
flavor conservation condition is
\beql{GCscons}
\langle N_{\rm s}\rangle  - \langle N_{\rm \bar s}\rangle=0\,,
\eeq
where the average is over the ensemble of physical systems, which in Gibbs 
sense are weakly connected, and can exchange particle number. Thus, each individual
system does not conserve strangeness, the fluctuations of strange and 
antistrange quark number, in the subsystem, are 
independent of each other. In each subsystem,
the magnitude of the average violation is the fluctuation in particle number:
\beql{GCscons1}
N_{\rm s}  -  N_{\rm \bar s}
\simeq  \sqrt{N_{\rm s}+N_{\rm \bar s}} \,.
\eeq

In heavy ion reactions where each collision system is completely disconnected
from the other, use of grand canonical method is an idealization which
allows the violation of the strangeness conservation law
 in the theoretical description of each individual collision reaction. 
This is a severe defect of the statistical method applied, which needs to 
be quantitatively understood and corrected. Only in a very large 
system, the average  yield of strange quarks nearly 
equals the average yield of antistrange quarks, and the relative violation
of strangeness conservation vanishes like $1/\sqrt{N_{\rm s}}$. 

For many reaction systems of physical interest, the difference 
in strangeness and antistrangeness 
yield is not negligible. We thus must improve the statistical description 
enforcing exact strangeness conservation both  for systems small and large.
Strangeness is always produced in pairs and all experiments always will
find (in absence of flavor changing weak interactions) that the micro canonical 
condition is satisfied,
\beql{MCscons}
N_{\rm s}  -  N_{\rm \bar s} =0 \,.
\eeq

The yield of net strangeness will vanish exactly
within our theoretical approach, as it does  in nature. We will
next show how it is possible  to implement that the net strangeness 
conservation law is satisfied exactly in the statistical description of
the physical properties, while using the power and convenience of statistical 
mechanics. This then has the minor defect that the
number of pairs, 
\beql{pairs}
\langle N_{\rm s}\rangle  +\langle N_{\rm \bar s}\rangle
         =2\langle N_{\mbox{\scriptsize s-pair}}\rangle\,,
\eeq
fluctuates in each   collision. This may even not be a defect
at all as the quantum mechanical laws which govern particle production 
also are leading to such fluctuations.  

We refer to this situation,
with exact conservation of some quantum number implemented,
here specifically strangeness, as the canonical statistical ensemble.
Each member of the ensemble conserves net strangeness exactly, while
the number of pairs fluctuates, being exchanged between the members of the
ensemble. 
The discussion above was for the case of vanishing strangeness
quantum number, but could be easily repeated
for the case of another arbitrary net value of the conserved quantum number. 

\subsection{Grand canonical and canonical partition functions}
\lssec{canon}
The grand partition function in the classical Boltzmann limit for strange particles 
has the form,
 \beqarl{Zstrange}
   \ln{\cal Z}_{\mathrm s}^{\mathrm{HG}} \equiv Z^{(1)}_{\rm HGs}=
    { {VT^3} \over {2\pi^2}} &\hspace{-0.3cm}\Bigl[&\hspace{-0.3cm}
     (\lambda_{\mathrm s} \lambda_{\mathrm q}^{-1}
 +
     \lambda_{\mathrm s}^{-1} \lambda_{\mathrm q}) \gamma_{\rm s}\gamma_{\rm q} F_{\mathrm K}
\nonumber 
 +
     (\lambda_{\mathrm s} \lambda_{\mathrm q}^{2} +
     \lambda_{\mathrm s}^{-1} \lambda_{\mathrm q}^{-2}) \gamma_{\rm s} \gamma_{\rm q}^2 F_{\mathrm Y}
\nonumber \\&\hspace{-0.3cm}&\hspace{-1.3cm}
 +
   (\lambda_{\mathrm s}^2 \lambda_{\mathrm q} + \lambda_{\mathrm s}^{-2}
     \lambda_{\mathrm q}^{-1}) \gamma_{\rm s}^2\gamma_{\rm q} F_\Xi
 +
    (\lambda_{\mathrm s}^{3} + \lambda_{\mathrm s}^{-3})
     \gamma_{\rm s}^3 F_\Omega \,     \Bigr] \, .
 \eeqar
In the phase space function
$F_i$, all kaon (K), hyperon (Y), cascade ($\Xi$) and omega
($\Omega$) resonances plus their antiparticles are taken into
account:
 \beqar
   F_{\mathrm K} &&\hspace{-0.6cm}=  \sum_j g_{{\mathrm K}_j} W(m_{{\mathrm K}_j}/T);\
           {\mathrm K}_j={\mathrm K},{\mathrm K}^\ast,{\mathrm K}_2^\ast,\ldots ,\ \
           m\le 1780 \ {\mathrm{MeV}}\, ,
   \nonumber\\
   F_{\mathrm Y} &&\hspace{-0.6cm}=  \sum_j g_{\mathrm Y_{\!j}} W(m_{\mathrm Y_{\!j}}/T);\
           \mathrm Y_{\!j}=\Lambda, \Sigma, \Sigma(1385),\ldots ,\ 
           m\le 1940\ {\mathrm{MeV}}\, ,
   \nonumber\\
   F_\Xi &&\hspace{-0.6cm}=  \sum_j g_{\Xi_j} W(m_{\Xi_j}/T);\
           \Xi_j=\Xi,\Xi(1530),\ldots ,\ \
           m\le 1950\ {\mathrm{MeV}}\, ,
   \nonumber\\
   F_\Omega &&\hspace{-0.6cm}=  \sum_j g_{\Omega_j}
W(m_{\Omega_j}/T);\            \Omega_j=\Omega,\Omega(2250)\, .
 \eeqarl{4b}
The $g_i$ are the spin--isospin degeneracy factors, $W(x)=x^2K_2(x)$,
where $K_2$ is the modified Bessel function. 

The chemical fugacities,
as introduced in \req{Zstrange}, allow to count separately the quark content 
($ \lambda_{\mathrm q}, \lambda_{\mathrm s})$ and the yield of 
quark--antiquark pairs ($ \gamma_{\rm s}, \gamma_{\rm q} $). Specifically,
 \beqarl{ns}
\langle N_{\rm s}  \rangle - \langle N_{\rm \bar s} \rangle
     &&\hspace{-0.6cm}= 
       \lambda_{\mathrm s}{\partial\over \partial \lambda_{\mathrm s}}
                           \ln{\cal Z}_{\mathrm s}^{\mathrm{HG}} \, ,\\
\langle N_{\rm s}  \rangle + \langle N_{\rm \bar s} \rangle
     &&\hspace{-0.6cm}=
    2\langle N_{\mbox{\scriptsize s-pair}}\rangle =  
        \gamma_{\mathrm s}{\partial\over \partial \gamma_{\mathrm s}}
                           \ln{\cal Z}_{\mathrm s}^{\mathrm{HG}} \,.
 \eeqarl{nspair}

To emphasize that any flavor (in particular s, c) or even baryon number
is under consideration here, we generalize slightly the notation $\rm s\to f$. 
We also expand the  exponential of the  one particle partition function $Z^{(1)}$ in \req{Zstrange}:
\beql{ZGCscl}
{\cal Z}_{\mbox{\scriptsize f}}=e^{Z^{(1)}_{\rm f}}
=
\sum_{n=0}^\infty\frac{1}{n!}\left(Z^{(1)}_{\rm f}\right)^n\,.
\eeq
The flavor and antiflavor terms within $Z^{(1)}_{\rm f}$ 
are additive in \req{Zstrange}, and we consider at first only 
singly-flavored particles,
 in a self explanatory simplified notation: 
\beql{Z1f}
Z^{(1)}_{\rm f}=
\gamma[\lambda_{\rm f} \tilde{F}_{\rm f} +\lambda_{\rm f}^{-1}\tilde{F}_{\rm \bar f}]\,,
\qquad\qquad \tilde{F}_i= \frac{VT^3}{2\pi^2}F_i.
\eeq
Combining \req{Z1f} with \req{ZGCscl}, we obtain:
\beql{ZGCsclexp}
{\cal Z}_{\rm f}=
\sum_{n,k=0}^\infty {{\gamma^{n+k}}\over {n! k!}}\lambda_{\rm f}^{n-k} 
\tilde{F}_{\rm f}^n\tilde{F}_{\rm \bar f}^k\,. 
\eeq

When $n\ne k$, the sum in \req{ZGCsclexp} contains contributions with unequal
number of $\rm f$ and $\rm \bar f$ terms. Only when $n=k$, we have 
contributions with exactly equal number of  $\rm f$ and $\rm \bar f$ terms.
We recognize that only $n=k$  terms contribute  to
the canonical partition function with exactly conserved flavor quantum number,
\beql{ZCsclexp}
{Z}_{\rm f=0}=
\sum_{n=0}^\infty {{\gamma^{2n}}\over {n! n!}}(\tilde{F}_{\rm f} \tilde{F}_{\rm \bar f})^n
 = {I}_0(2\gamma \sqrt{\tilde{F}_{\rm f} \tilde{F}_{\rm \bar f}})\,,
\eeq
where we have introduced the modified Bessel 
function $I_0$.

The argument  of $I_0$ 
has a physical meaning, it is  the yield of 
flavor pairs $N_{\rm pair}^{\rm GC}$ in grand canonical ensemble,
evaluated  with 
grand canonical flavor conservation, \req{GCscons}. To see this,  we evaluate:
\beql{fGC}
0=\frac{\partial}{\partial \lambda_{\rm f}}\ln {\cal Z}_{\rm f}
=
\frac{\partial}{\partial \lambda_{\rm f}}\left(\gamma[\lambda_{\rm f} \tilde{F}_{\rm f} 
                 +\lambda_{\rm f}^{-1}\tilde{F}_{\rm \bar f}]\right) \,.
\eeq
We obtain: 
\beql{scond}
\left.\lambda_{\rm f}\right|_0=\sqrt{\tilde{F}_{\rm \bar f}/\tilde{F}_{\rm f}}\,,
\qquad\left.\ln {\cal Z}_{\rm f}\right|_{\lambda_{\rm f}=\lambda_{\rm f}|_0}
=
2\gamma \sqrt{\tilde{F}_{\rm f} \tilde{F}_{\rm \bar f}}
\equiv 
2\,N_{\rm pair}^{\rm GC}
\,.
\eeq

In order to evaluate, using  \req{ZCsclexp},
the number of flavor pairs in the canonical ensemble, 
we need to average the number $n$ over
all the contributions to the sum in \req{ZCsclexp}. To obtain the extra factor $n$, we 
perform the differentiation with respect to $\gamma^2$ and obtain the canonical ensemble 
f-pair yield,
\beqarl{Cspairs}
\langle N_{\mathrm f}^{\mathrm{CE}} \rangle \equiv
\gamma^2\frac{d}{d\gamma^2}\ln {Z}_{\rm f=0}
=
\gamma \sqrt{\tilde{F}_{\rm f} \tilde{F}_{\rm\bar f}} 
\frac{{I}_1(2\gamma \sqrt{\tilde{F}_{\rm f} \tilde{F}_{\rm\bar f}})}
        {{I}_0(2\gamma \sqrt{\tilde{F}_{\rm f} \tilde{F}_{\rm\bar f}})}
=N_{\rm pair}^{\rm GC}
       \frac{{I}_1(2N_{\rm pair}^{\rm GC})}{{I}_0(2N_{\rm pair}^{\rm GC})} \,,
\eeqar
where we have used ${I}_1(x)=d{I}_0(x)/dx$. The first term is identical with
the result we obtained in the grand canonical
formulation, \req{scond}. The second factor $I_1/I_0$ 
is the effect of exact conservation of the number of 
flavor pairs.

\subsection{Projection method}
\lssec{proj}
For the case of `Abelian' quantum numbers, \eg, flavor or baryon 
number, the projection method arises from the general relation
between the grand canonical and canonical partition function:
\beql{RelGCC}
{\cal Z}(\beta,\lambda,V)=
  \sum_{f=-\infty}^{\infty}\lambda^{\rm f} Z_{\rm f}(\beta,V)\,.
\eeq
In the canonical partition function  $Z_{\rm f}$, some discrete
(flavor, baryon) quantum number has the value f. 
Substituting $\lambda=e^{i\varphi}$, we obtain: 
\beql{Zcanphi}
Z_{\rm f}(\beta,V;N_{\rm f})=\int_0^{2\pi} \frac{d\varphi}{2\pi}e^{-iN_{\rm f} \varphi}{\cal Z}(\beta,\lambda
            =e^{i\varphi},V)\,.
\eeq

In case of  Boltzmann limit, and including singly charged
particles only, we obtain for net flavor $N_{\rm f}$, from \req{ZGCsclexp}:
\beql{expCan}
Z_{\rm f}(\beta,V;N_{\rm f})=\sum_{n,k=0}^\infty {{\gamma^{n+k}}\over {n! k!}}
  \int_0^{2\pi} \frac{d\varphi}{2\pi}e^{i(n-k-N_{\rm f})\varphi} \tilde{F}_{\rm f}^n\tilde{F}_{\rm\bar f}^k\,.
\eeq
The integration over $\varphi$ yields the $\delta(n-k-N_{\rm f})$-function. Replacing $n=k+N_{\rm f}$, we obtain:
\beql{expCanfs}
Z_{\rm f}(\beta,V;N_{\rm f})=\sum_{k=0}^\infty {{\gamma^{2k+N_{\rm f}}}\over {k! (k+N_{\rm f})!}}
   \tilde{F}_{\rm f}^{k+N_{\rm f}}\tilde{F}_{\rm\bar f}^k\,.
\eeq
The power series definition of the modified Bessel function $I_{\rm f}$  is:
\beql{Ifdef}
I_{N_{\rm f}}(z)=\sum_{k=0}^\infty \frac{(z/2)^{2k+N_{\rm f}} }{k!(k+N_{\rm f})!}\,.
\eeq
Thus, we obtain:
\beql{Canf}
Z_{\rm f}(\beta,V;N_{\rm f})=\left(\frac{\tilde{F}_{\rm f}}{\tilde{F}_{\rm\bar f}}\right)^{N_{\rm f}/2} 
 \!\!\!\!\!\!        I_{N_{\rm f}}(2\gamma \sqrt{\tilde{F}_{\rm f} \tilde{F}_{\rm\bar f}})\,.
\eeq
The case  $N_{\rm f}=0$, we considered earlier \req{ZCsclexp}, is reproduced.
We note that for integer $N_{\rm f}$, we have $I_{N_{\rm f}}=I_{-N_{\rm f}}$.
We used $N_{\rm f}$ as we would count baryon number, thus in flavor counting, 
$N_{\rm f}$ counts the flavored quark content, with quarks counted positively and antiquarks negatively. 
This remark is relevant in numerical studies 
when the factors ${\tilde{F}_{\rm f}},\,\tilde{F}_{\rm\bar f}$ 
contain baryochemical potential. 

\subsection{Suppression of multistrange particle yield}
\lssec{multi}
Multistrange particles can be
introduced as additive terms in the exponent of \req {Zcanphi}.
This allows us to evaluate their yields \bcite{Ham00c}. 
However, the canonical partition function
is dominated by singly strange particles and we will 
assume, in the following, that it 
is sufficient to only consider these, in order to obtain the 
effect of canonical flavor conservation.
This assumption is consistent with use of classical Boltzmann statistics. In fact,
expanding the Bose distribution for kaons, one finds that the next to leading
order contribution, which behaves as 
strangeness $N_{\rm s}=\pm2$ hadron, is dominating in the projection
the influence of all multistrange hadrons. 

In order to find yields of rarely produced  particles
 such as is, \eg, $\Omega(\rm sss)$,
we show the omega term explicitly:
\beql{OmeeffZC}
Z_{\rm f}(\beta,V;N_{\rm f}=0)=\int_0^{2\pi} \frac{d\varphi}{2\pi}
        e^{\tilde{F}_{\rm f}e^{i\varphi} + \tilde{F}_{\rm\bar f}e^{-i\varphi}
            +\lambda_\Omega e^{3i\varphi}\tilde{F}_\Omega+\cdots}\,.
\eeq
The unstated terms in the exponent are the other small 
abundance multi-flavored particles.  
The fugacities not associated with strangeness, 
as well as the yield fugacity
$\gamma_{\rm s}$, are incorporated in \req{OmeeffZC} 
into the phase space factors $\tilde{F}_i$ for simplicity 
of notation. 

The number of  $\Omega$ is obtained differentiating 
$\ln Z_{\rm f}(\beta,V)$, with respect to $\lambda_\Omega$,
 and subsequently neglecting  the sub dominant terms in 
the exponent. We obtain: 
\beql{OYCan}
\langle N_\Omega\rangle=\frac{\tilde{F}_\Omega}{I_0} 
    \int_0^{2\pi} \frac{d\varphi}{2\pi}e^{3i\varphi}
        e^{\tilde{F}_{\rm f}e^{i\varphi} + \tilde{F}_{\rm\bar f}e^{-i\varphi}}\,.
\eeq
The integral is just $Z_{\rm f}(\beta,V;N_{\rm f}=-3)$, 
\req{Canf}, since we need to balance the three 
strange quarks in the particle observed by the balance 
in the background of singly
strange particles (kaons and hyperons):
\beql{OYCanFin}
\langle N_\Omega\rangle=\tilde{F}_\Omega \left(\frac{\tilde{F}_{\rm f}}{\tilde{F}_{\rm\bar f}}\right)^{-3/2}  
\frac{I_3 (2N_{\rm pair}^{\rm GC})}{I_0 (2N_{\rm pair}^{\rm GC})}\,.
\eeq
We recall that, according to \req{scond}, 
the middle term is just the fugacity factor $\lambda_{\rm s}^3$.
The first two factors, in \req{OYCanFin}, 
constitute the grand canonical yield, while the canonical $\Omega$-suppression
factor is the last term. A full treatment of the canonical suppression of 
multistrange particles in small volumes has been used to obtain particle yields in 
elementary interactions \bcite{Bec98b}.

Similarly, one finds that the $\Xi$  suppression has the factor $I_2/I_0$,
while as discussed for the general example of flavor pair yield, 
the single strange particle yield is suppressed by the factor $I_1/I_0$.
The yield of all flavored hadrons in the canonical approach (superscript `C') 
can  be written as function of the yield 
expected in the grand canonical approach 
in the general form,
\beql{yieldShad}
\langle s^\kappa\rangle^{\rm C}=
        \tilde{F}_\kappa \left(\frac{\tilde{F}_{\rm f}}{\tilde{F}_{\rm\bar f}}\right)^{\kappa/2} 
           \frac{I_{|\kappa|} (2N_{\rm pair}^{\rm GC}) }{I_0 (2N_{\rm pair}^{\rm GC}) }
=
\langle s^\kappa\rangle^{\rm GC}\frac{I_{|\kappa|} (2N_{\rm pair}^{\rm GC}) }
                             {I_0  (2N_{\rm pair}^{\rm GC}) } \,,
\eeq
with $\kappa=\pm3,\,\pm2,$ and $\pm1$ for $\Omega,\ \Xi$, 
and Y, K, respectively. On the left hand
side, in \req{yieldShad}, the power indicates the 
flavor content in the particle considered
with negative numbers counting antiquarks. 
We note, inspecting the final form of 
\req{yieldShad}, that the canonical suppression 
of particles and antiparticles is the same. 
However, a particle/antiparticle  asymmetry can 
occur if baryon/antibaryon asymmetry is present.

The simplicity of this result originates in
the assumption that the single strange particle contribution to strangeness 
conservation are dominant. 
A more complex evaluation taking all multistrange hadrons into account, 
but considering kaons as Boltzmann particles
is theoretically inconsistent.

\section{Canonical strangeness and charm suppression}      
\lsec{CanSup}
\subsection{The suppression function}      
\lssec{suppress}
The canonical flavor yield suppression factor, 
\beql{defeta}
\eta\equiv\frac{{I}_1(2\gamma \sqrt{\tilde{F}_{\rm f} \tilde{F}_{\rm\bar f}})}
       {{ I}_0(2\gamma \sqrt{\tilde{F}_{\rm f} \tilde{F}_{\rm\bar f}})}
=\frac{{I}_1(2N_{\rm pair}^{\rm GC})}{{I}_0(2N_{\rm pair}^{\rm GC})} < 1\,,
\eeq
depends in a complex way on the volume of the system, or alternatively said,
on the grand canonical number of pairs, $N_{\rm pair}^{\rm GC}$. The suppression function 
$\eta(N)\equiv I_1(2N)/I_0(2N)$ is shown in \rf{PLQUETA}, as function of $N$. For $N>1$, we see (dotted lines) 
that the approach to the grand  canonical limit is relatively slow, it follows the asymptotic form, 
\beql{etaasym}
\eta\simeq 1-\frac{1}{4 N}-\frac{1}{128 N^2}+\ldots \,,
\eeq
while for  $N \ll 1 $, we see a nearly linear rise:
\beql{eta0lim}
\eta={N}-\frac{N^3}{2}+\ldots\,
\eeq
Overall, when the yield of particles is small, we have using \req{eta0lim}:
\beql{nGClim}
N_{\mathrm{f}}^{\rm CE}\simeq  (N_{\mathrm f}^{\rm GC})^2\,.
\eeq

\myfig{0}{12}{PLQUETAF}{0}{-4.1}{-1}{tb}
\capt{PLQUETA}{Canonical suppression function of N}{
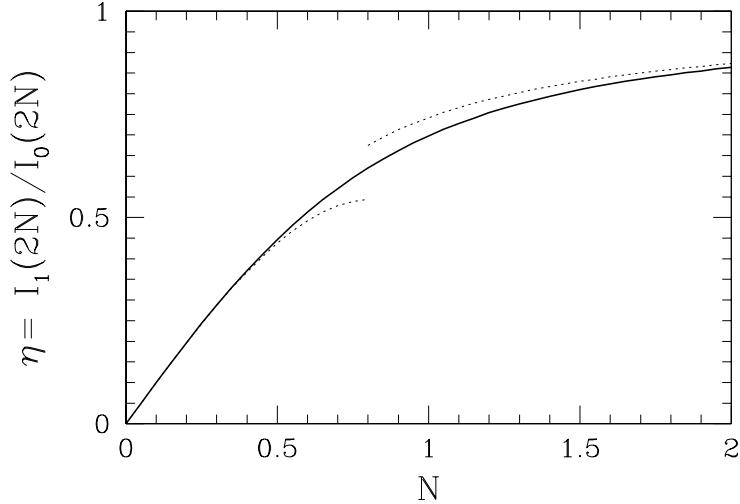
Solid line: canonical yield suppression factor as function of the grand canonical 
particle yield $N$. Dotted lines: asymptotic expansion presented in text.
 }
\end{figure}

The chemical equilibrium yield, at small abundances,
is quadratic in grand canonical particle yield, which
 for $m>T$ is, expanding the $K_2$-Bessel function,
\beql{NGClim}
N_{\mathrm{f}}^{\rm GC} =  
 \frac{g_{\rm f}}{2\pi^2}T^3V \sqrt{\frac{\pi m_{\rm f}^3}{2T^3}}e^{-m_{\rm f}/T}\,.
\eeq
Thus, when the yield of particles is small, \eg, when $m_{\rm f}\gg T$, 
the canonical result applies:
\beql{nCanlim}
N_{\rm f}^{\rm CE}=\frac{g_{\rm f}^2}{4\pi^3}  T^3m_{\rm f}^3 V^2  e^{-2m_{\rm f}/T}\,.
\eeq
This result resolves an old puzzle first made explicit 
by Hagedorn, who queried  the quadratic behavior of the pair 
particle yield, compared to Boltzmann yield,
$Y \propto e^{-2m/T} \simeq ( e^{-m/T})^2$
being concerned about rarely occurring 
astrophysical pair production processes \bcite{Hag71}.

The benchmark result, seen
in \rf{PLQUETA}, is that
when  one particle pair  would be expected to be present in
grand canonical chemical equilibrium  the 
actual canonical yield is suppressed, 
the true phase space yield is 0.6 pairs. 
This suppression occurs when the exact strangeness conservation
is enforced due to reduction of the accessible phase space by
particle--antiparticle correlation.

\myfig{0}{12}{PLQUETA123F}{0}{-4.1}{-1}{tb}
\capt{PLQUETA123}{Canonical suppression 3s 2s as function of N}{
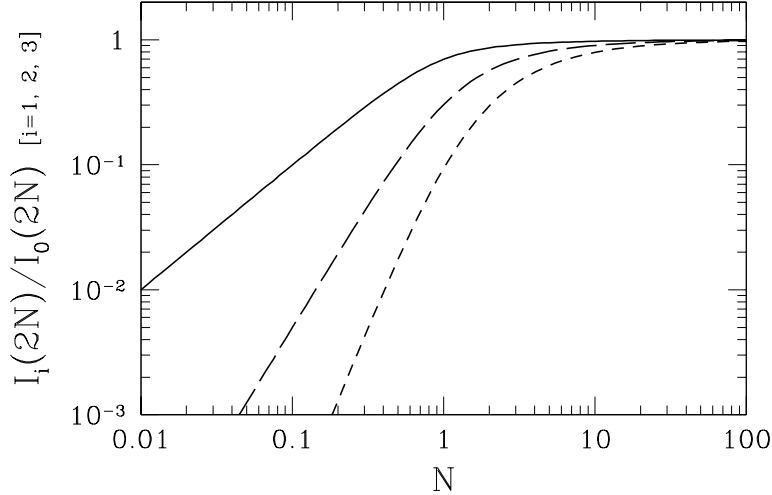
Canonical yield suppression factor $I_\kappa/I_0$ 
as function of the grand canonical 
particle yield $N$. Short-dashed line: suppression of triply strange 
hadrons; long dashed:  suppression of doubly flavored hadrons; and 
solid line, the suppression of singly flavored hadrons.
\index{canonical!ensemble!particle  suppression} }
\end{figure}
We now look at the suppression of multistrange particles 
by the suppression factors $\eta_3(N)=I_3(2N)/I_0(2N)$, for $\Omega$,
and 
$\eta_2(N)=I_2(2N)/I_0(2N)$,
for $\Xi$. For small values of $N$, we obtain: 
\beql{etaiNs}
\eta_\kappa\equiv \frac{I_\kappa(2N)}{I_0(2N)}\to N^\kappa
  \frac1{\kappa!}\left(1-\frac{\kappa}{\kappa+1} N^2\right)\,.
\eeq
This result is easily understood on physical grounds: for example when
the expected grand canonical yield is three strangeness pairs, it is 
quite rare that all three strange quarks go into an $\Omega$. This 
is seen in \rf{PLQUETA123} (short dashed curve), and in fact  this
will occur 1/5 as often as we would expect computing the yield of 
$\Omega$, ignoring the canonical conservation of strangeness.
The other lines, in \rf{PLQUETA123},  correspond to the other suppression 
factors, long dashed is $\eta_2(N)=I_2(2N)/I_0(2N)$ 
and the solid line is $\eta(N)=I_1(2N)/I_0(2N)$.
They are shown dependent on the 
number $N$ of strange pairs expected in the 
grand canonical equilibrium.
We see that the suppression effect increases with 
strangeness content, and
that for $N>5$, it practically vanishes.

\subsection{Hadronic gas compared to quark--gluon plasma}   
\lssec{HGQGP}   
We first consider how big a volume we need, in order to 
find (using grand canonical ensemble counting) one pair of strange particles.
As unit volume, we choose  $V_{\rm h}=(4\pi/3)\, 1$\,fm$^3$.  
The flavor and antiflavor phase space is 
symmetric in the deconfined state. In the  Boltzmann limit,
\beql{Fsbaars}
\tilde{F}_{\rm f}=\tilde{F}_{\rm\bar f}
=\frac{3VTm^2_{\rm f}}{\pi^2}K_2(m_{\rm f}/T)\,.
\eeq
In \rf{PLRR0}, the dashed line shows the volume required for one pair
 using the strange quark phase space, 
which does not depend on $\lambda_{\rm q}$, and has been 
obtained choosing $m_{\rm s}=160$ MeV and  $T=160$ MeV. 
Just a little less than  one hadronic volume 
suffices,  one finds one pair in $V_{\rm h}$ for $m_{\rm s}=200$ MeV. 

\myfig{0}{12}{PLRR0F}{0}{-4.1}{-1}{tb}
\capt{PLRR0}{Volume for one strange pair in GC ensemble}{
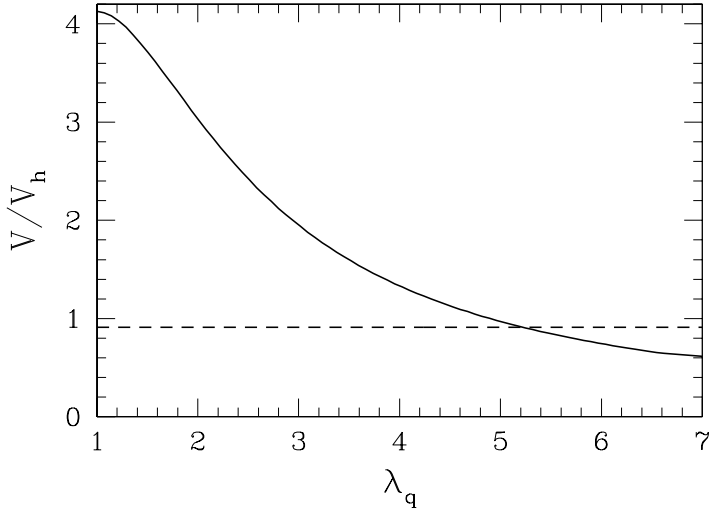
Volume needed  for one strange quark pair using grand canonical counting
as function of $\lambda_{\rm q}$ for $T=160$ MeV,
 $\gamma_{\rm q}=1, \gamma_{\rm s}=1$,
 $V_{\rm h}=(4\pi/3)\, 1$\,fm$^3$. 
Solid line: hadron gas phase space, dashed line:
quark phase space with $m_{\rm s}=160$ MeV.}
\end{figure}

For the hadronic phase space,  counting as before strange quark content as positively 
`flavor charged', we obtain using  \reqa{Zstrange}{4b}:
\beqarl{Fexpl}
\tilde{F}_{\rm f}&\hspace{-0.3cm} =&\hspace{-0.3cm}
    \lambda_{\mathrm q}^{-1} \tilde{F}_{\mathrm K} +\lambda_{\mathrm q}^{2}  \tilde{F}_{\mathrm Y}\,,\\
\tilde{F}_{\rm\bar f}&\hspace{-0.3cm} =&\hspace{-0.3cm}
   \lambda_{\mathrm q}  \tilde{F}_{\mathrm K}+\lambda_{\mathrm q}^{-2}  \tilde{F}_{\mathrm Y}\,.
\eeqarl{Fbarexpl}
All these quantities $\tilde{F}_i$ are proportional to the reaction volume. 
With $\lambda_{\rm s}$ chosen to conserve strangeness,
\req{scond},
\beql{VV0had}
\frac{V}{V_{\rm h}}= \frac{2\pi^2}{
V_{\rm h}T^3\gamma_{\rm q}\gamma_{\rm s}\sqrt{(F_{\rm K}+\lambda_{\rm q}^3F_{\rm Y})
                                             (F_{\rm K}+\lambda_{\rm q}^{-3}F_{\rm Y})}
}\,.
\eeq
The result is shown as solid line in \rf{PLRR0},
as function of $\lambda_{\rm q}$, for $\gamma_{\rm q}=1, \gamma_{\rm s}=1$.
We recall that at SPS and RHIC energies, we have $\lambda_{\rm q}<1.6$. 
We see that for small $\lambda_{\rm q}$, we need much greater volumes to find
one strange quark pair, and thus we recognize that the hadron gas phase space 
is significantly smaller in absence of dense baryon 
number. In a more colloquial language,
strangeness `production' is easier in the channel 
$\Lambda\rm\overline K$ than in $\rm K\overline K$.

This large difference in the magnitude of the 
phase space between the confined and deconfined phase,
seen in \rf{PLRR0}, makes the 
effect of canonical suppression different when
we compare quark--gluon plasma with hadronic gas. Thus in what follows
the yield of strange hadrons is dependent on the nature 
of the phase from which emission occurs. 

It has been  proposed to exploit the canonical suppression,
which grows with strangeness content, in order to explain the increase
of strange hadron production, which also 
grows with strangeness content of the particle \bcite{Ham00c,Red01}.
To do this, we must turn things `upside down' by rebasing all yields 
to unity at a unit volume. We first consider more
closely how big an effect we get for singly strange
hadrons for quark--gluon plasma and hadronic gas. In \rf{PLQUENCH2}, 
the quark phase (solid line) and hadron phase (dashed line), the
suppression results are renormalized multiplicatively to cross for $V=V_{\rm h}$ 
unity. Since quark phase space  is bigger, it has `less space left'
to grow to reach saturation, and hence the production enhancement is by
a factor two, while for  the hadron case there is  `more catch up left' to do 
and thus the enhancement is larger, we see that it is by a factor three.

\myfig{0}{12}{PLQUENCH2F}{0}{-4.1}{-1}{tb}
\capt{PLQUENCH2}{Canonical suppression function of V}{
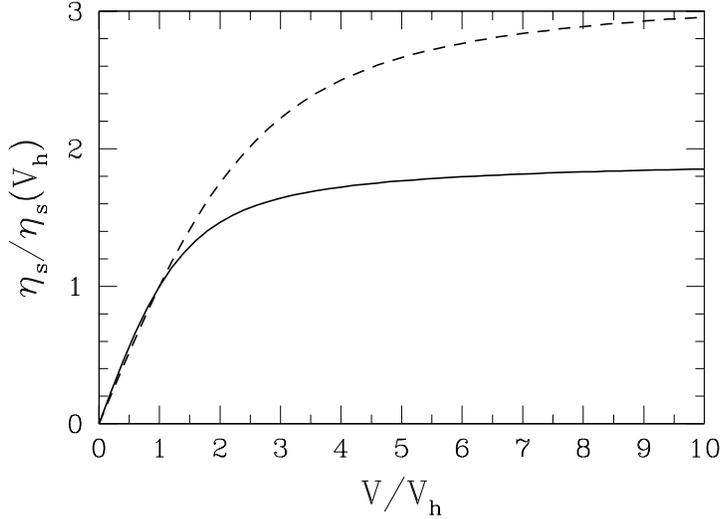
Canonical yield enhancement at large volumes compared to unit 
hadron volume  $V_{\rm h}=(4\pi/3)\, 1$\,fm$^3$. 
Solid line QGP phase, dashed line HG. 
\index{canonical!ensemble!particle enhancement} }
\end{figure}

\subsection{Canonical charm yields}      
\lssec{charm}
Not everybody is tempted to use statistical equilibrium 
when considering the yield of charm. The charmed 
quark mass is sufficiently high to stop even the greatest of 
optimists from claiming that thermal collisions could equilibrate the
yield. On the other hand, since the mass is so large, the thermal 
grand canonical abundance is relatively small. Thus, the few hard collisions
occurring between colliding partons also suffice to reproduce
so much charm that it can easily be well above the chemical 
equilibrium yield. 

The yield of charm, in Pb--Pb interactions at 158$A$ GeV, is 
estimated from lepton background at 0.5 pairs per central collision \bcite{Abr01a}.
We  can  use the small $N$ expansion, \req{etaiNs}. 
The corresponding A--A canonical enhancement factor, 
compared to  p--A, is $N_{\mathrm{AA}}/N_{\mathrm{pA}}\simeq 100$ $A$. (Here, $N$ is 
now grand canonical yield of `open' charm, and not strangeness). 
Experimental results are  scaling with $A^{\alpha}\,,\ \alpha<1.3$, thus
there is no space for canonical enhancement/suppression for 
charm production of this magnitude. 

To be more specific, we show, in 
\rf{PLQUEYIELDCH2}, the specific yield per unit volume as function of volume
of charm $\langle N\rangle_{\rm pair}$. The canonical effect is the
deviation from a constant value and it is significant,  ${\cal O}(100)$. 
Even at $V=400V_{\rm h}$ the  infinite volume
grand canonical limit is not yet attained,  for the 
case of the larger phase space of QGP (solid line), the total charm yield is 
0.8 charm  pairs. The absolute yield in both phases is 
strongly dependent on temperature used, here $T=145$ MeV,
corresponding to SPS hadronization condition. In 
quark--gluon plasma, we took  $m_{\rm c}=1.3$ GeV.
The hadronic gas phase space includes all known charmed mesons and baryons,
with light quark abundance controlled by
$\mu_{\mathrm{b}}=210$ MeV, $\mu_{\rm s}=0$.

\myfig{0}{12}{PLQUEYIELDCH2F}{0}{-4.1}{-1}{tb}
\capt{PLQUEYIELDCH2}{Charm specific canonical yield as function of V}{
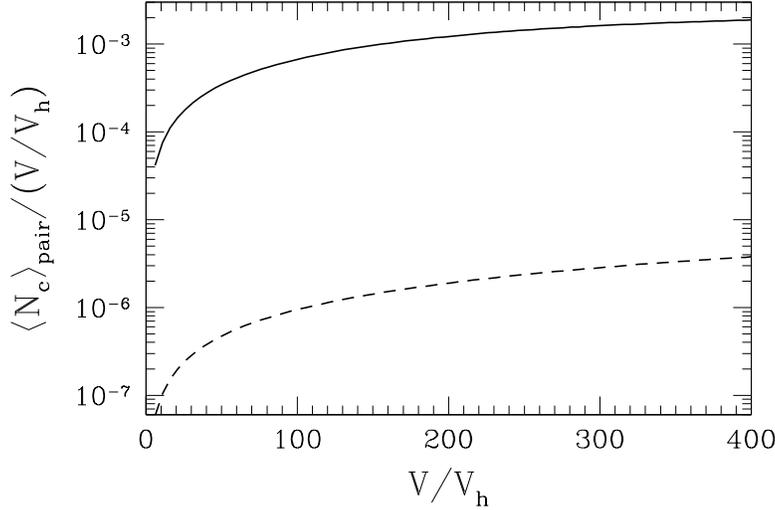
Canonical yield of open charm quark pairs $\langle N\rangle_{\rm pair}$
per unit volume as function of volume, in units of $V_{\rm h}=4\pi/3\, 1$\,fm$^3$.
Solid line: QGP with $m_{\rm c}=1.3$ GeV, dashed line HG at $\mu_{\mathrm{b}}=210$ MeV, 
both phases at $T=145$ MeV. 
\index{charm!canonical suppression} }
\end{figure}

While choosing a slightly higher value of $T$, we could increase the
equilibrium yield of charm in hadronic gas to the quark--gluon 
plasma level \bcite{Gor01}, 
this does not eliminate the effect of
canonical suppression of charm production if chemical 
equilibrium is assumed for charm in the elementary interactions. 
We are simply so deep in the `quadratic' domain of the yield,
see \req{etaiNs}, that playing with parameters changes nothing, since
we are constrained in Pb--Pb interactions by experiment 
to have a charm yield below 
one pair. Then, the expected yield in p--p and p--A interactions 
is well below measurement, the canonical suppression is
overwhelming. Charm yield  is surely not in chemical equilibrium 
either at small or large volumes, most probably in both limits.

\section{Final remarks}      \lsec{close}
We have  discussed the subtle differences in particle yields that arise in 
equilibrium statistical mechanics when, within a finite system, 
the conservation of flavor is enforced exactly. We addressed
the recent proposal \bcite{Ham00c,Red01},  that the enhancement of strange
particles may be also described in chemical equilibrium model.

Compared to the grand canonical ensemble, we see, 
in \rf{PLQUETA123},   `upside down' suppression/enhancement factor 
which depends sensitively on the choice of the (grand-canonical) yield
 of strange pairs $N\propto V $.  Thus, with an appropriate choice 
of a reference point $V_{\rm h}$ and  $T$ these  factors can be  
fine tuned as is in fact
done in Ref.\,\bcite{Ham00c,Red01}, within a eyeball fit.
For the purpose of the following discussion  it is important to remember
that the experimental enhancement results are reported by the
WA97 experiment, with base obtained in p--Pb and p--Be collision system \bcite{And99}.

For p--p reactions
at the top SPS energy the strange pair yield is believed to be 
$\langle N_{\mbox{\scriptsize s-pair}}\rangle = 0.66\pm0.07$ 
\bcite{Wro85}. However, since the reference experiment for
the enhancement has been p--Be and more generally p--A \bcite{Red01}, 
we  consider in  the bottom section of \rf{PLQUETA123XF} twice as 
large strangeness yield.  To obtain \rf{PLQUETA123XF} we have taken 
the results  shown in \rf{PLQUETA123}, converted the ordinate to be 
the canonical yield,  $N^{\rm CE}=NI_1(2N)/I_0(2N)$, and normalize 
the yields at the reference  pair yields $0.66\pm0.07$ and $1.3\pm0.2$, 
thus showing the `canonical enhancement',  $E_i,\ i=1,2,3$, with reference  
to the p--p and (as estimate) to the  p--Be collision system 
in \rf{PLQUETA123XF}.

\myfig{2.5}{13}{PLQUETA123XD}{0}{-1.5}{-2}{tb}
\capt{PLQUETA123XF}{Canonical enhancement as function of NCE}{
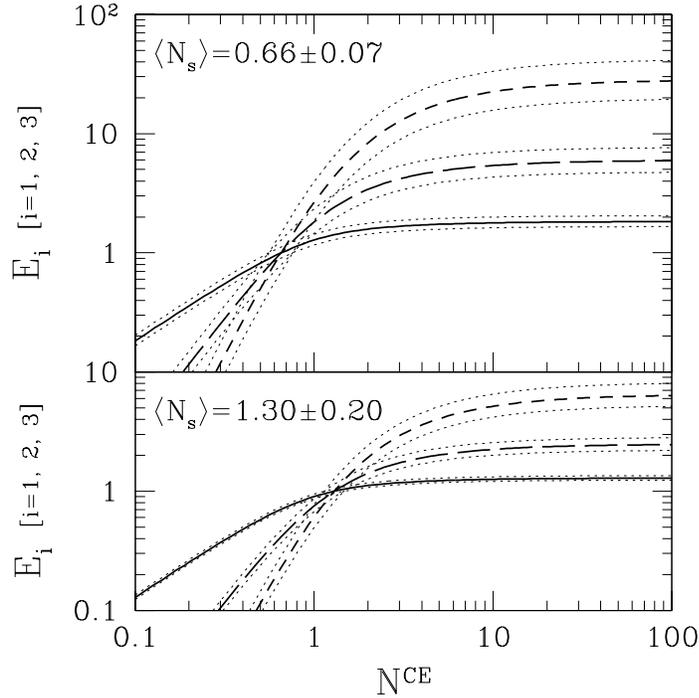
Canonical yield enhancement factor $E_i,\ i=1,2,3$
as function of the  canonical pair
particle yield $N^{CE}$. Solid line, $E_1$ the enhancement of 
singly flavored hadrons, relative to the 
physical (canonical) yields $\langle N_{\rm s}\rangle=0.66\pm0.07$,
(top section) and $\langle N_{\rm s}\rangle =1.3\pm0.2$ (bottom section)
expected  in p--p and p--Be reactions, 
respectively. Similarly,  long dashed:  $E_2$ 
enhancement of doubly flavored hadrons; and 
short-dashed line: $E_3$ enhancement of triply strange 
hadrons. Dotted lines correspond to the errors arising from 
the error in the strangeness yield,
to which the results are normalized. 
} 
\end{figure}

The single strange hadron enhancement, see the
solid line in \rf{PLQUETA123XF}, is by factor 1.2--2. Had 
we taken for the reference the p--Pb reactions with 
as much as 3--4 times larger total
strangeness  yield as in the p--p system, the canonical enhancement 
would have largely disappeared. This is a result of the rather rapid 
disappearance of the canonical suppression as function of reaction
volume \bcite{Raf80}, as is seen in \rf{PLQUENCH2}. The grand canonical
yields are  reached in a few elementary collision 
volumes, or equivalent when a  few strange quark pairs are present. 

On the other hand, the experimental results from 
NA52 experiment \bcite{Kab99b} show a rather sudden 
strangeness enhancement 
threshold at $\simeq 50$ participants, just where NA57 
recently reports a sudden onset of
$\overline\Xi$ yield enhancement \bcite{Eli01}. While the results
of NA57 are still being reconciled in significant detail with its
predecessor WA97,
there is no disagreement regarding the sudden onset of strangeness
production. 

This threshold behavior is easily understood:
the yield rise can be significantly delayed 
in medium sized system reflecting on the fact that kinetic 
processes need to be invoked to establish
chemical equilibrium, which in fact is perhaps only attained  
at 30--50 elementary  volumes, where new physics comes 
into play.  The shape of the enhancement curve 
as function of the volume also indicates where the new (gluon fusion)
mechanism  of strangeness production sets in \bcite{Raf82b},
\ie\ where one would expect that
the deconfinement begins, as function of reaction volume at given 
collision energy.

We see in \rf{PLQUETA123XF} the long-dashed line 
describing the double strange (cascade) canonical enhancement which 
is by  a factor $E_2=$2.2--5.5, and 
the short-dashed line describing the enhancement of triply strange $\Omega, 
\overline\Omega$ by a factor $E_3=$6--20. We recognize, also when
 comparing to the experimental data that:\\
i) the canonical enhancement of multistrange baryons 
is very sensitive to the reference system considered,\\
ii) the canonical enhancement
occurs already for very small systems and its primary variation 
is realized in p--p, p--Be, p--Pb, where it is not observed;\\
iii) the multistrange hadron enhancement appears in experiment 
for systems much larger than expected from canonical 
equilibrium considerations.

In passing, we address the more complex case of 
$\phi(\rm s\bar s)$ enhancement. 
This particle has only `hidden' strangeness, 
it does not follow the $E_2$ enhancement curve. 
If the  $\phi$ production mechanism in p--p and A--A reactions 
are the same and specifically arise from $\rm s$--$\rm \bar s$ pair
yields in the  fireball, than the enhancement of $\phi$  production 
follows the general strangeness pair enhancement. On the basis of 
canonical mechanism we would expect less than two fold 
$\phi(\rm s\bar s)$ enhancement, while experimentally
$\phi(\rm s\bar s)$ enhancement  is by factor 3.6  \bcite{Afa00}, 
comparing p--p with Pb--Pb. 

For charm, we have  obtained a canonical  enhancement well above 
experimental expectations. We have seen, in \rf{PLQUEYIELDCH2}, 
that a large change is expected 
in the  canonical charm yield per unit of volume 
(which is equivalent to yield per participant) when chemical equilibrium
is subsumed. Experimental
results do not show that the yield of charm 
is rising that fast. This implies that  heavy charm quarks are
not in chemical equilibrium, and their production has to be studied in kinetic
theory of parton collision processes.

If attainment of 
chemical equilibration is seen as a fundamental process driven by an 
unknown `demon' which operates within statistical hadronization,
charm should not be different from strangeness. Thus,  if charm
is excluded from equilibrium, this  means that there is indeed
no 21st century Maxwell `equilibration demon' control of charm, 
and  by extension,  also not of strangeness. 

In conclusion, we  have shown that the canonical 
strangeness enhancement:\\ 
1) lacks internal theoretical consistency,
considering both strangeness and charm;\\
2) disagrees with  the strangeness yield enhancement seen in 
experiment as function of reaction volume;\\
3) effect for multistrange hadrons is highly sensitive 
to the properties of the reference system, and hence the 
`explanation' of experimental 
results in \bcite{Ham00c,Red01} is at best coincidental; \\
4)  behavior as function of volume (\ie\ strangeness yield) 
 for p--p, p--Be, p--Pb disagrees with the available 
experimental results.

This work has demonstrated that the chemical equilibrium
canonical suppression/enhancement reinterpretation 
of quark--gluon plasma strange hadron signature is 
not able to explain the full set of experimental results available today. 

\subsection*{Acknowledgments}
One of us (JR) would like to thank Krzysztof Redlich for 
valuable comments. 
Work supported in part by a grant from the U.S. Department of
Energy,  DE-FG03-95ER40937. Laboratoire de Physique Th\'eorique 
et Hautes Energies, University Paris 6 and 7, is supported 
by CNRS as Unit\'e Mixte de Recherche, UMR7589.



\subsection*{References}
\begin{thebibliography}{99}

\bibitem{Mag55}
V.B. Magalinskii and Ia.P. Terletskii, 1955.
 The statistics of charge-conserving systems and its application to
the theory of multiple production.
{\em J. Exper.Theoret. Phys. USSR} {\bf 29}, 151. Translated: 1956.
{\em Sov. Physics. JETP} {\bf 2}, 143.

\bibitem{Shu74}
E.V. Shuryak, 1974.
Final-state interactions and the composition of secondary particles.
{\em Yad. Fiz.} {\bf 20}, 549. Translated: 1975. 
{\em Sov. J. Nuc. Phys} {\bf 20}, 295.

\bibitem{Raf80}
J.~Rafelski and M.~Danos, 1980.
 The importance of the reaction volume in hadronic collisions.
 {\em Phys. Lett. {\rm B}}, {\bf 97}, 279.

\bibitem{Ham00c}
S.~Hamieh, K.~Redlich, and A.~Tounsi, 2000.
 Canonical description of strangeness enhancement from {p--A} to
  {Pb--Pb} collisions.
 {\em Phys. Lett {\rm B}}, {\bf 486}, 61.

\bibitem{Red01}
K.~Redlich, S.~Hamieh, and A.~Tounsi, 2001.
 Statistical hadronization and strangeness enhancement from {p--A} to
  {Pb--Pb} collisions.
 {\em J. Phys. {\rm G}}, {\bf 27}, 413.

\bibitem{Red80}
K.~Redlich and L.~Turko, 1980.
 Phase transitions in the statistical bootstrap model with an internal
  symmetry.
 {\em Z. Physik {\rm C}}, {\bf 5}, 201.

\bibitem{Tur81}
L.~Turko, 1981.
 Quantum gases with internal symmetry.
 {\em Phys. Lett. {\rm B}}, {\bf 104}, 153.

\bibitem{Cle91}
J.~Cleymans, K.~Redlich, and E.~Suhonen, 1991.
 Canonical description of strangeness conservation and particle
  production.
 {\em Z. Physik {\rm C}}, {\bf 51}, 137.

\bibitem{Der85}
C.~Derreth, W.~Greiner, H.-Th. Elze, and J.~Rafelski, 1985.
 Strangeness abundances in $\rm \overline{p}$--{nucleus}
  annihilations.
 {\em Phys. Rev. {\rm C}}, {\bf 31}, 1360.

\bibitem{Bec98}
F.~Becattini, M.~Ga\'zdzicki, and J.~Sollfrank, 1998.
 On chemical equilibrium in nuclear collisions.
 {\em Eur. Phys. J. {\rm C}}, {\bf 5}, 143.

\bibitem{Elz83}
H.-Th. Elze, W.~Greiner, and J.~Rafelski, 1983.
 On the color-singlet quark--gluon plasma.
 {\em Phys. Lett. {\rm B}}, {\bf 124}, 515.

\bibitem{Elz84}
H.-Th. Elze, W.~Greiner, and J.~Rafelski, 1984.
 Color degrees of freedom in a quark--gluon plasma at finite baryon
  density.
 {\em Z. Physik {\rm C}}, {\bf 24}, 361.


\bibitem{Raf82b}
J.~Rafelski and B.~{M\"uller}, 1982.
 Strangeness production in the quark--gluon plasma.
 {\em Phys. Rev. Lett.}, {\bf 48}, 1066.
 See: {\it Phys. Rev. Lett.}, {\bf 56}, 2334E (1986).

\bibitem{Let00b}
J.~Letessier and J.~Rafelski, 2000.
Observing quark--gluon plasma with strange hadrons.
{\em Int. J. Mod. Phys. {\rm E}}, {\bf 9}, 107.


\bibitem{Bec98b}
F.~Becattini.
 Universality of thermal hadron production in $\rm pp$, $\rm p\bar p$
  and $\rm e^+e^-$ collisions.
 In L.~Cifarelli, A.~Kaidalov, and V.A. Khoze., editors, {\em
  Universality Features in Multihadron Production and the Leading Effect}.
  World Scientific, Singapore, 1998.

\bibitem{Hag71}
R.~Hagedorn.
 {\em Lectures on thermodynamics of strong interactions}.
 CERN-Yellow Report 71-12, 1971.

\bibitem{Wro85}
A.~Wr{\'o}blewski, 1985.
 On the strange quark suppression factor in high energy collisions.
 {\em Acta Phys. Pol. {\rm B}}, {\bf 16}, 379.

\bibitem{Abr01a}
M.C. {Abreu {\it et al.}, NA50 collaboration}, 2001.
 Results on open charm from {NA}50.
 {\em J. Phys. {\rm G}}, {\bf 27}, 677.

\bibitem{Gor01}
 M.I.~Gorenstein, H.~{St\"o}cker, A.P.~Kostyuk and W.~Greiner, 2001.
 Statistical coalescence model with exact charm conservation.
 {\em Phys. Lett. {\rm B}}, {\bf 509}, 277.



\bibitem{And99}
E. Andersen {\it et al.}, WA97  collaboration, 1999,
Strangeness enhancement at mid-rapidity in Pb--Pb collisions at 158\agev
{\em Phys. Lett. {\rm B}}, {\bf 449}, 401. For latest results
see http://wa97.web.cern.ch/WA97/.

\bibitem{Kab99b}
S.~{Kabana {\it et al.}, NA52 collaboration}, 1999.
 Centrality dependence of $\pi^{\pm}$, {K}$^{\pm}$, baryon and
  antibaryon production in {Pb}+{Pb} collisions at 158\agev.
 {\em J. Phys. {\rm G} Nucl. Part. Phys.}, {\bf 25}, 217.

\bibitem{Eli01}
D.~{Elia {\it et al.}, NA57 collaboration}, 2001.
 Results on cascade production in {Pb--Pb} interactions from the
  {NA}57 experiment.
 {\em hep-ex/0105049}.

\bibitem{Afa00}
S.V. Afanasev {\it et al.}, NA49 collaboration, 2000,
Production of $\phi$ mesons in p+p, p+Pb and central Pb+Pb 
collisions at $E_{\mbox{beam}}$ 158$A$ GeV.
{\em Phys. Lett. {\rm B}}, {\bf 491}, 59.


\end{thebibliography}
\end{document}